\documentclass[twocolumn,aps,prl,preprintnumbers,bibnotes10pt,superscriptaddress]{revtex4}

\usepackage{graphicx}
\usepackage{dcolumn}
\usepackage{epsfig}
\usepackage{bm}
\usepackage{array}
\usepackage{float}
\usepackage{amsmath}
\usepackage{lipsum}
\usepackage{color}
\usepackage{bbm}

\definecolor{MainRed}{RGB}{69, 187, 52}


\begin{document}
\title{Self-bound quantum droplets in atomic mixtures}
\author{G. Semeghini}
\affiliation{LENS and Dipartimento di Fisica e Astronomia, Universit\'a di Firenze, 50019 Sesto Fiorentino, Italy} 
\affiliation{CNR Istituto Nazionale Ottica, 50019 Sesto Fiorentino, Italy}
\author{G. Ferioli}
\affiliation{LENS and Dipartimento di Fisica e Astronomia, Universit\'a di Firenze, 50019 Sesto Fiorentino, Italy}
\author{L. Masi} 
\affiliation{LENS and Dipartimento di Fisica e Astronomia, Universit\'a di Firenze, 50019 Sesto Fiorentino, Italy} 
\author{C. Mazzinghi} 
\affiliation{LENS and Dipartimento di Fisica e Astronomia, Universit\'a di Firenze, 50019 Sesto Fiorentino, Italy} 
\author{L. Wolswijk} 
\affiliation{LENS and Dipartimento di Fisica e Astronomia, Universit\'a di Firenze, 50019 Sesto Fiorentino, Italy} 
\author{F. Minardi} 
\affiliation{CNR Istituto Nazionale Ottica, 50019 Sesto Fiorentino, Italy}
\affiliation{LENS and Dipartimento di Fisica e Astronomia, Universit\'a di Firenze, 50019 Sesto Fiorentino, Italy}
\author{M. Modugno}
\affiliation{Depto. de F\'isica Te\'orica e Hist. de la Ciencia, Universidad del Pais Vasco UPV/EHU, 48080 Bilbao, Spain}
\affiliation{IKERBASQUE, Basque Foundation for Science, 48011 Bilbao, Spain}
\author{G. Modugno}
\affiliation{LENS and Dipartimento di Fisica e Astronomia, Universit\'a di Firenze, 50019 Sesto Fiorentino, Italy}
\affiliation{CNR Istituto Nazionale Ottica, 50019 Sesto Fiorentino, Italy}
\author{M. Inguscio}
\affiliation{CNR Istituto Nazionale Ottica, 50019 Sesto Fiorentino, Italy}
\affiliation{LENS and Dipartimento di Fisica e Astronomia, Universit\'a di Firenze, 50019 Sesto Fiorentino, Italy}
\author{M. Fattori} 
\affiliation{LENS and Dipartimento di Fisica e Astronomia, Universit\'a di Firenze, 50019 Sesto Fiorentino, Italy}
\affiliation{CNR Istituto Nazionale Ottica, 50019 Sesto Fiorentino, Italy}

\date{\today}

\begin{abstract}
Self-bound quantum droplets are a newly discovered phase in the context of ultracold atoms. In this work we report their experimental realization following the original proposal by Petrov [Phys. Rev. Lett. {\bf 115}, 155302 (2015)], using an attractive bosonic mixture. In this system spherical droplets form due to the balance of competing attractive and repulsive forces, provided by the mean-field energy close to the collapse threshold and the first-order correction due to quantum fluctuations. Thanks to an optical levitating potential with negligible residual confinement we observe self-bound droplets in free space and we characterize the conditions for their formation as well as their equilibrium properties. This work sets the stage for future studies on quantum droplets, from the measurement of their peculiar excitation spectrum, to the exploration of their superfluid nature.     
\end{abstract}

\maketitle

Ultracold atoms are commonly known and studied in their gas-phase. They are confined on a finite volume by external potentials, but they readily expand as they are released from their container. A recent theoretical proposal \cite{Petrov} has surprisingly pointed out that a Bose-Bose mixture of diluted weakly-interacting atomic gases can form liquid-like droplets, which are self-bound in free space and whose equilibrium densities are independent on the atom number. When the attraction between the two atomic species becomes larger than the single-species average repulsion, the mixture is expected to collapse according to mean-field (MF) theory \cite{Cornell}. In this regime, instead, an effective repulsion provided by the first beyond-mean-field correction to the energy, the so-called Lee-Huang-Yang (LHY) term \cite{LHY}, arises to arrest collapse and stabilize the system. The equilibrium between the two competing forces leads to the formation of a self-bound droplet, while the isotropic nature of the van der Waals interactions shapes its spherical geometry. This new quantum state of matter is expected to display a number of interesting features. The most peculiar among them is related to its excitation spectrum. In a specific region of the droplet phase diagram, the particle emission threshold is predicted to lie below any possible excitation mode \cite{Petrov}. Any excess of energy is thus expelled by losing particles, leading to an effective self evaporation and keeping the droplet at zero temperature.

\begin{figure*}[ht] 
\begin{center}
\includegraphics {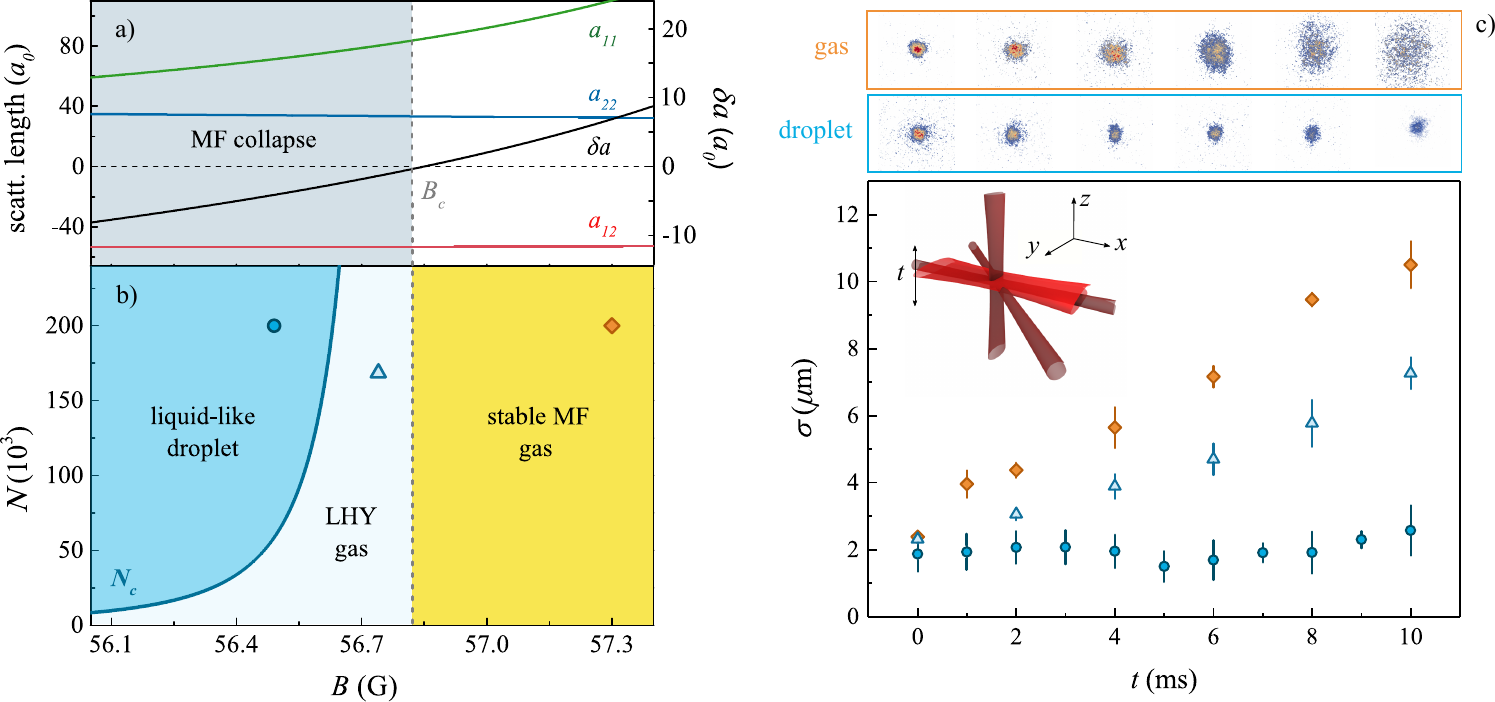}
\end{center}
\caption{a) Intra and inter-species scattering lengths between the hyperfine states $|1, 0\rangle$ (state 1) and $|1, -1\rangle$ (state 2)  of $^{39}$K, tuned by an external magnetic field $B$ via Feshbach resonances. The resulting MF energy of the mixture is proportional to the effective scattering length $\delta a$, which becomes negative at $B_c=56.85$ G. b) Phase diagram for the mixture as a function of the atom number $N$ and of the magnetic field $B$. c) Evolution of the cloud in free space for three different points of the phase diagram in b). The upper rows show the difference between the evolution of the density profiles in the gas and droplet phases. Inset: Schematic representation of the geometry of the experiment.}  \label{phaseDiagram}
\end{figure*}

Very recently the stabilization mechanism generated by the LHY correction has been recognized responsible for the formation of a different class of self-bound quantum systems, i.e. dipolar droplets \cite{Pfau1,Pfau2,Pfau3,Ferlaino, Pfau4, Santos, Blakie1}. 
While attractive mixtures create spherical droplets, in dipolar gases droplets are elongated along the dipole direction and thus strongly anisotropic. The different geometry, together with the different kind of interactions governing the stabilization, lead to important differences in the properties of the two objects and enriches the range of phenomena that can be explored. A primary example of this concerns the excitation spectrum, where self evaporation is expected to occur only in the mixture case, being strongly related to the droplet isotropic shape \cite{Blakie2}.

In this work we study experimentally the formation of spherical quantum droplets in a homonuclear bosonic mixture. Exploiting magnetic Feshbach resonances we tune the interatomic scattering lengths to reach the interaction regime where the mixture is predicted to be self-bound. We implement an optical levitating potential with negligible residual confinement along all directions, which allows to have long interrogation times and access the droplet properties in free space. We probe the mixture phase-diagram, proving the existence of a self-bound phase and identifying the critical conditions for its formation. We characterize the equilibrium properties of the droplet and we find a good agreement with the theoretical predictions in \cite{Petrov}. Finally, we discuss the dynamics observed in the droplet formation and compare it to the result of a numerical simulation.  

We create self-bound droplets using two hyperfine states of $^{39}$K, namely $|F=1, m_F=0\rangle$ (state 1) and $|F=1, m_F=-1\rangle$ (state 2). Feshbach resonances allow to tune the mutual contact interactions as represented in Fig.\ref{phaseDiagram}a as a function of the magnetic field $B$ \cite{Efimov}. The intra-species scattering lengths $a_{11}$ and $a_{22}$ are both positive, while the inter-species $a_{12}$ is negative. We define an effective scattering length for the mixture $\delta a=a_{12}+\sqrt{a_{11} a_{22}}$, which becomes negative for $B<B_c$, setting the threshold for collapse in the usual MF picture \cite{Cornell}. The stabilization effect of the LHY correction predicted in \cite{Petrov} appears exactly here. Contrary to the case of a single species \cite{dalfovo}, in a mixture of BECs the MF and LHY terms have a different dependence on the interparticle scattering lengths. While the MF energy, $E_{MF}$, is proportional to $|\delta a|$ and thus vanishes close to $B_c$, the LHY correction, $E_{LHY}$, scales with the average of $a_{11}$ and $a_{22}$, thus becoming comparable to the MF term in this regime. Moreover, the two terms have a different dependence on the density $n$, since $E_{MF}\propto n^2$ while $E_{LHY}\propto n^{5/2}$. This means that when the MF contribution becomes negative leading to an uncontrolled increase of density and eventually to collapse, the positive LHY term, having a steeper dependence on $n$, arrests the collapse and stabilizes the system. In this regime the mixture can be found in two different phases depending on the total atom number $N$. When $N$ is larger than a critical number $N_c$ \cite{Petrov}, the mixture forms a self-bound liquid-like droplet. Below that threshold, the kinetic energy overcomes the MF attraction and the system goes back into an expanding gas phase, labeled as LHY gas in the phase diagram of Fig.\ref{phaseDiagram}b.

We prepare a Bose-Einstein condensate (BEC) of $^{39}$K atoms in state 2, in a crossed dipole trap, created by three red-detuned laser beams, with trapping frequencies $\omega_{x}=2\pi \times 250(10)$ Hz, $\omega_{y}=2\pi \times 240(10)$ Hz and $\omega_{z}=2\pi \times 280(10)$ Hz, along the axes sketched in Fig.\ref{phaseDiagram}c. A homogeneous magnetic field is used to tune the scattering lengths as in Fig.\ref{phaseDiagram}a. Starting with a BEC with up to $4\times10^5$ atoms, we ramp linearly the magnetic field in 20 ms to a desired target value and then we apply a radio-frequency pulse of 10 $\mu$s to transfer $\sim 50\%$ of the atoms in state 1. In order to observe the subsequent evolution for sufficiently long times, remaining within the field of view of our imaging system, gravity compensation is required. 
The vertical position of a red-detuned elliptical laser beam is modulated in time with an acousto-optical modulator at a frequency of 4 kHz, such that the averaged potential experienced by the atoms provides a gradient opposite to gravity (red beam in the inset of Fig.\ref{phaseDiagram}c). A large waist on the horizontal direction ($y$) and a suitable time-modulation along the vertical direction ($z$) guarantee negligible residual curvatures on all directions \cite{SI}. At the end of the RF pulse, we switch off the dipole traps and switch on the levitating potential to observe the expansion of the mixture. After a variable waiting time we record the density profile of the cloud via absorption imaging along the $y$ direction \cite{SI}. We fit it with a two-dimensional gaussian and measure the size along $x$ and $z$ as the half width at $1/e^2$.  
In order to probe the different phases expected for the mixture, we study the expansion of the cloud in three different points of the phase diagram of Fig.\ref{phaseDiagram}b. In Fig.\ref{phaseDiagram}c we report the average size $\sigma=\sqrt{\sigma_x \sigma_z}$ as a function of the expansion time $t$. The mixtures prepared with $\delta a>0$ (orange diamonds) or $\delta a<0$ and $N<N_c$ (light blue triangles) show the typical gas behavior: when released from the dipole trap they expand at a finite velocity. For $\delta a<0$ and $N>N_c$ (blue circles), instead, the size of the cloud remains constant for 10 ms, proving the formation of a self-bound droplet. 

\begin{figure}[h!] 
\begin{center}
\includegraphics {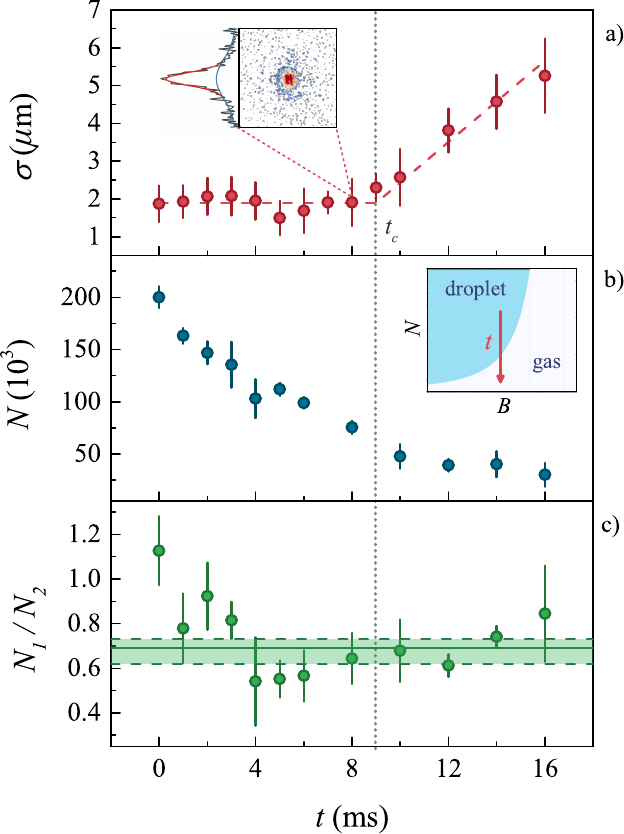}
\end{center}
\caption{Time evolution of $\sigma$ (a), $N$ (b) and $N_1/N_2$ (c) in the droplet phase at $B=56.45(1)$ G. The inset in a) reports the density profile of the droplet close to $t_c$, together with the fitted bimodal function. In the inset in b) we draw a sketch of the trajectory followed by the system in the mixture phase-diagram during the time evolution, due to losses. The dashed line in a) represents the heuristic fit described in the text and used to identify the critical time $t_c$. In c) the solid line represents the theoretical equilibrium value $N_1/N_2=\sqrt{a_{22}/a_{11}}$ and the green area between the dashed lines includes the allowed deviations $\delta N_i/N_i \sim |\delta a|/a_{ii}$. The error bars represent the statistical uncertainty and correspond to one standard deviation.}  \label{8.97all}
\end{figure}

In order to characterize the droplet phase, we also perform measurements of the total atom number and of the relative population in state 1 and 2. After a variable expansion time, we perform a Stern-Gerlach separation of the two components, by applying a magnetic field gradient along the vertical direction $z$, so that we can count separately $N_1$ and $N_2$. In Fig.\ref{8.97all} we report the evolution of the size $\sigma$, measured as in Fig.\ref{phaseDiagram}c, of the total atom number $N=N_1+N_2$ and of the ratio $N_1/N_2$, for $B=56.45(1)$ G. The data in Fig.\ref{8.97all}a show that the mixture is in the self-bound regime only up to a critical time $t_c$, while afterwards it expands as a gas. We estimate the value of $t_c$ by fitting our data with a heuristic piecewise function composed by a flat line at short times, plus a linear growth after $t_c$ (dashed line in Fig.\ref{8.97all}a). The behavior of $\sigma(t)$ is explained by the evolution of $N(t)$ in Fig.\ref{8.97all}b. The atom number drops quite rapidly in the first 10 ms, so that the system follows the phase-diagram trajectory sketched in the inset of Fig.\ref{8.97all}b: at a given expansion time the atom number goes below $N_c=N(t_c)$ and the system undergoes a droplet-to-gas transition. In order to understand the loss dynamics observed in Fig.\ref{8.97all}b, we have to consider two different effects. The first one is three-body recombination, which causes strong losses, mainly in state 1 \cite{SI}. The second one is related to a stabilization mechanism of the mixture: the droplet forms with a specific population imbalance, $N_1/N_2=\sqrt{a_{22}/a_{11}}$ \cite{Petrov}. It can only bear a deviation from that value which corresponds to an increase of the atom number in one of the two components $\delta N_i/N_i \sim |\delta a|/a_{ii}$, $i=1,2$. Any excess of atoms beyond this threshold is not bound to the droplet and expands as a gas. Combining these two mechanisms, we can interpret the behavior observed in Fig.\ref{8.97all}b-c. In the first 4 ms the ratio $N_1/N_2$ decreases, meaning that the mixture is mainly losing atoms in state 1. This is due both to the stronger three-body losses in that component and to the initial ratio $N_1/N_2\sim 1$ being significantly larger than its equilibrium value (green area in Fig.\ref{8.97all}c). The droplet thus expels the exceeding atoms in state 1, which expand away from it and get effectively lost during the Stern-Gerlach sequence. When the mixture reaches the equilibrium population imbalance, the ratio $N_1/N_2$ stabilizes, meaning that three-body losses in state 1 are accompanied by a release of atoms in state 2 into the unbound fraction. This is also compatible with the observed bimodal density profiles in the measurements of Fig.\ref{8.97all}a, with a dense central part constant in size, corresponding to the droplet, surrounded by low density tails corresponding to the unbound atoms. We fit the profiles with the sum of two gaussians and associate the size of the droplet to the width of the central one (inset of Fig.\ref{8.97all}a).  

\begin{figure}[h]
\begin{center}
\includegraphics {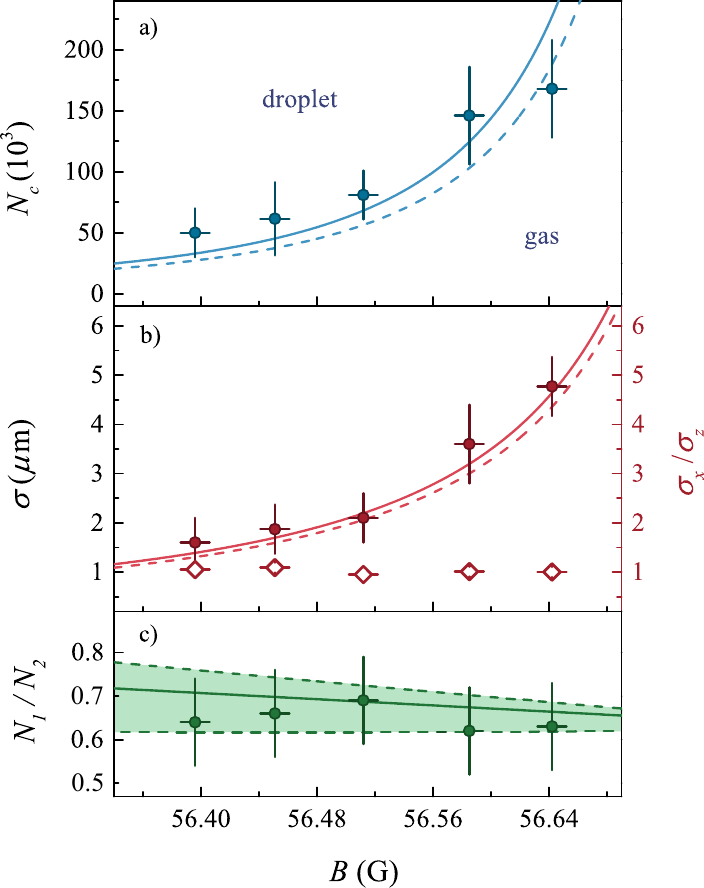}
\end{center}
\caption{Critical atom number $N_c$ (a) and equilibrium properties $\sigma$ (b) and $N_1/N_2$ (c) of the droplet as a function of the magnetic field $B$. In b) we also report the aspect ratio $\sigma_x/\sigma_z$ (diamonds). The curves in a) and b) correspond to the theoretical predictions for the metastable (dashed) and stable (solid) self-bound solutions \cite{Petrov}. In c) the theoretical curves are obtained as in Fig.\ref{8.97all}c, considering the equilibrium value of $N_1/N_2$ (solid line) and the allowed deviations (green area). The vertical error bars include the uncertainty coming from the determination of $t_c$ and the statistical uncertainty. The horizontal ones are due to the uncertainty in the magnetic field calibration. All error bars correspond to one standard deviation.}  \label{Nc}
\end{figure}

We repeat the measurements of Fig.\ref{8.97all} for different values of the magnetic field $B$. In Fig.\ref{Nc} we report the results for the critical number $N_c$ and for the size $\sigma$ and ratio $N_1/N_2$ also measured at $t_c$. The size reported in Fig.\ref{Nc}b corresponds to the geometrical average of the measured $\sigma_z$ and $\sigma_x$, plotted together with the aspect ratio $\sigma_x/\sigma_z$. We compare these experimental results with the theoretical predictions obtained from \cite{Petrov} and we find a good agreement within our error bars over the whole magnetic field range we explored.

\begin{figure}[h]
\begin{center}
\includegraphics {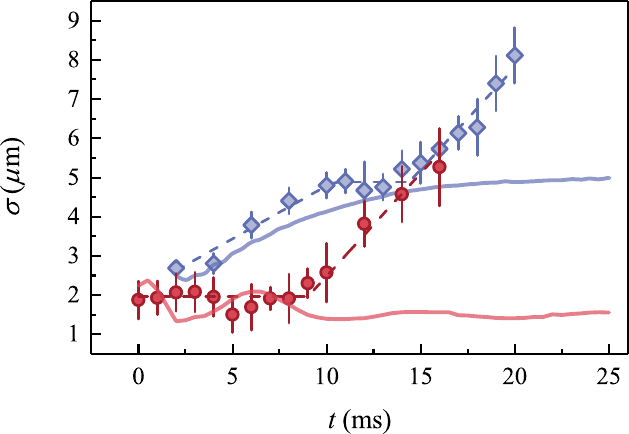}
\end{center}
\caption{Evolution of the size $\sigma$ for $B=56.45(1)$ G (circles) and $B=56.64(1)$ G (diamonds). We compare the experimental data with the result of the numerical simulations described in the text (solid lines). The dashed lines correspond to the heuristic fits performed to estimate the critical time $t_c$ in the two regimes.}  \label{equilibration}
\end{figure}

It is important to stress that the procedure we use to form the droplet starting from a single component BEC is a non-adiabatic process. Moreover, due to significant three-body losses in the system, the droplet has a limited lifetime. For this reason we decided to start from a BEC in an almost spherical confinement, with in-trap size close to the droplet one for $B<56.5$ G, i.e. $\sim 2$ $\mu$m. Minimizing the initial dynamics, we could observe the equilibrium configuration of the droplet, as visible from the measurements in Fig.\ref{8.97all}, where the size of the cloud is essentially constant up to $t_c$. However, for the two largest values of the magnetic field we considered, the evolution of $\sigma (t)$ is qualitatively different (Fig.\ref{equilibration}). In these cases the equilibrium size of the droplet, significantly larger than the BEC size, is reached after an initial expansion. After this transient, the size is constant for a given time and then it increases again when the atom number drops below $N_c$. In this case we perform a different heuristic fit to deduce the critical time $t_c$, that includes an additional linear growth of $\sigma$ at early times (blue dashed line in Fig.\ref{equilibration}).
In order to verify that the observation timescale is sufficient to reach equilibrium, we perform a simulation of the droplet dynamics in our experimental conditions \cite{SI}. We integrate numerically a system of two generalized Gross-Pitaevskii equations, which include first order quantum corrections in the local chemical potential \cite{Santos} via the two-species LHY term discussed in \cite{Petrov}. In Fig.\ref{equilibration} we compare the experimental data for $\sigma(t)$ at $B=56.45(1)$ G and $B=56.64(1)$ G with the result of the simulations. In order to capture the stabilization of the size at long times, three-body losses are not included in the numerical simulations. For the smallest value of $B$, the size slightly oscillates around its equilibrium value and finally stabilizes there, consistently with the plateau observed in the experiment. At large $B$, the equilibration time of the simulation is compatible with the initial transient of the experimental data, justifying the assumption that the central plateau in the evolution of $\sigma$ corresponds to the equilibrium size of the droplet.
The study of this dynamics allows to add a further consideration. For $B=56.64(1)$ G, even if the sample is prepared out of equilibrium, no oscillation of the size is visible, i.e. there is no signature of excitations. In the simulation this is associated with an expulsion of atoms from the droplet to the expanding low density tails. This is compatible with the predicted self evaporation, since discrete excitation modes are allowed only for $N>10^6$ at this magnetic field \cite{Petrov}. In our experiment it was difficult to distinguish a self-evaporation effect in the losses dynamics discussed above, but the observed stabilization of the size is a first hint of the occurrence of this phenomenon.

In conclusion, our work represents the first experimental observation of self-bound droplets in an atomic Bose-Bose mixture in free space. The measurement of the size and of the critical atom number as a function of the MF attraction provides a confirmation of the theoretical model proposed by Petrov \cite{Petrov}. The dynamics observed in the droplet formation indicates the main perspective of this work, which is the study of self evaporation, providing a quantitative evidence for it and probing how it disappears by increasing the atom number, or by adding an external confinement to induce an anisotropic geometry. Further studies may include the effect of a coherent coupling between the two atomic species in the droplet, exploring how its equilibrium properties are modified \cite{Salasnich}. It will also be interesting to investigate differences and analogies with dipolar droplets and different self-bound quantum fluids like helium clusters \cite{Stringari,HeReview}, probing the superfluid behavior \cite{Grebenev,Gomez} and comparing the excitation spectra \cite{Casas}. Finally our work triggers the possibility to investigate the formation of self-bound droplets in different atomic mixtures, possibly with reduced three body loss rates and longer lifetimes \cite{Ospelkaus,Modugno,Minardi,Arlt}.
\\

While completing this work, we became aware of a related work, reporting the observation of droplets in atomic mixtures in a one-dimensional lattice: C. R. Cabrera, L. Tanzi, J. Sanz, B. Naylor, P. Thomas, P. Cheiney, and L. Tarruell, arXiv:1708.07806 (2017).
\\

We acknowledge insightful discussions with D. Petrov, A. Simoni, S. Stringari, L. Tarruell, L. Tanzi and our colleagues of the Quantum Degenerate Gases group at LENS. We thank G. Spagnolli for assistance in the early stages of the experiment. This work was supported by the ERC Starting Grant AISENS No. 258325 and by EC-H2020 Grant QUIC No. 641122.
\\


\section{Supplemental Material}

\subsection{Optical levitating potential}

\begin{figure}[h!] 
\begin{center}
\includegraphics[width=0.55\columnwidth] {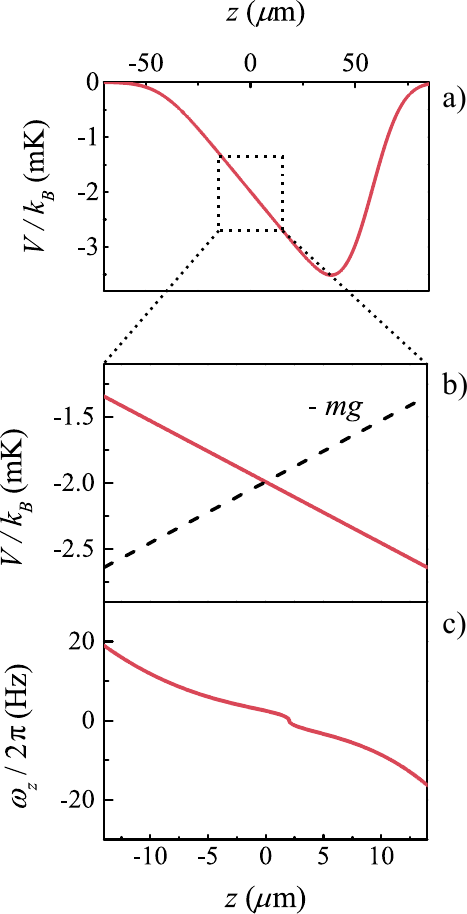}
\end{center}
\caption{a) Profile of the levitating potential $V(z)$. b) Profile of $V(z)$ around $z=0$ compared with gravity. c) Curvature produced by the levitating potential along the vertical direction.}  \label{RB}
\end{figure}

The optical potential used to levitate the atoms against gravity is created by a single elliptical beam, whose vertical position is modulated in time. The laser wavelength is $\lambda=1064$ nm and the waists of the beam along $z$ and $y$ (axes defined as in Fig.1c in the paper) are $w_z=23(3) \mu$m and $w_y=985(8) \mu$m. While the intensity of the beam is fixed, its vertical position $z_0$ changes periodically in time. The modulation function is defined on a single period $T$ as 

\begin{equation}
z_0(t)=A (2 \sqrt{|1-2 t/T|}-1)+C
\end{equation}  

where $A=50$ $\mu$m is the modulation amplitude and $C=7$ $\mu$m is an offset chosen to place the zero-curvature point at $z=0$, that corresponds to the position of the atoms. The beam is moved along the vertical direction by means of an acousto-optic modulator whose radio-frequency is modulated in time, thus changing the diffraction angle. The time-averaged potential along $z$ is then

\begin{equation}
V(z)=\frac{1}{T} \int_0^{T}{V_0 e^{-2(z-z_0(t))^2/w_z^2}},
\end{equation}
 
where $V_0$ is the potential amplitude. The profile of $V(z)$ is reported in Fig.\ref{RB}a). By adjusting the power in the laser beam to $\simeq 2.4$ W, the gradient produced in the central part of $V(z)$ is opposite to gravity (Fig.\ref{RB}b) and thus allows to levitate the atoms and access long observation times. We use a modulation frequency $\nu_{mod}=1/T=4$ kHz, which is fast enough that the atoms are only affected by the averaged potential. 

\begin{figure}[h!] 
\begin{center}
\includegraphics[width=0.9\columnwidth] {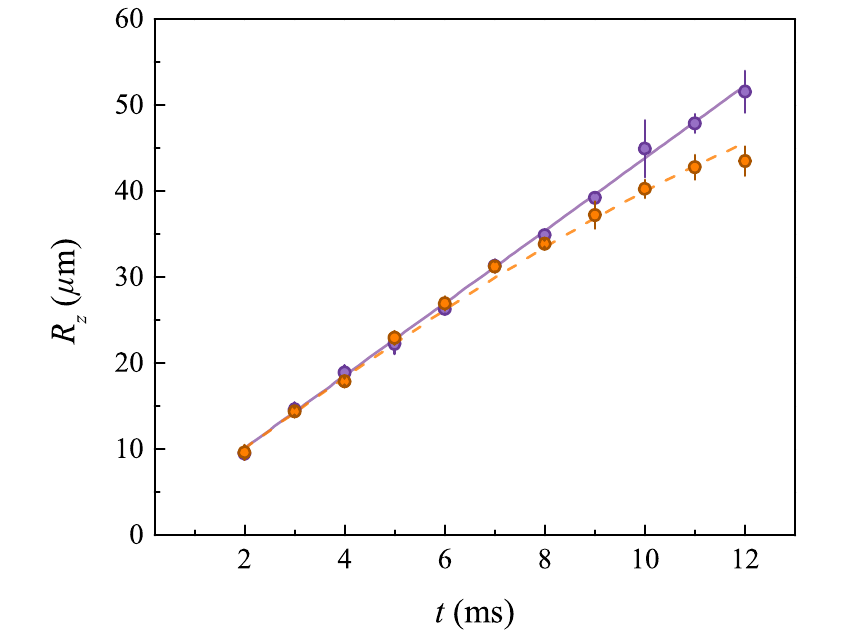}
\end{center}
\caption{Expansion of a BEC in the levitating potential vs free space. We measure the Thomas-Fermi radius along the vertical direction $R_z$ after releasing a BEC from its dipole traps. The purple points correspond to the free-space evolution, while the orange ones to the expansion in the levitating potential. The theoretical curves are the result of the evolution calculated from the Gross-Pitaevskii equation in free-space (solid purple) and with a vertical confinement $\omega_z=2\pi\times12$ Hz (dashed orange). }  \label{RBexp}
\end{figure}

The residual curvatures created by the levitating potential are weak along all directions. Along $z$, the calculated curvature is zero at $z=0$, is positive above and negative below, reaching $|\omega_z|\sim 2\pi \times 20$ Hz at a distance of $\sim 15 \mu$m from the center (Fig.\ref{RB}c). Along the other two directions, the curvature is positive but very weak: $\omega_x=2\pi\times2.2$ Hz and $\omega_y=2\pi\times7$ Hz. While the confinement along these two axes is well controlled by measuring the power and waists of the laser beam, we would like to perform an in-situ calibration of the curvature along $z$, which is more sensitive to misalignments. To perform a precise experimental measurement of $\omega_z$ is actually quite difficult, since the shape of the potential is far from being harmonic and the estimated curvatures are very weak. We perform two different measurements: in the first one we provide an upper bound to the local curvature in $z=0$, while in the second one we estimate the global effect produced by the levitating potential on a larger scale. The measurement of the local curvature is performed as follows: we measure the trapping frequency of a dipole trap aligned on the $z=0$ position, first in presence of a magnetic levitating potential ($\omega_{magn}$) and then with the optical levitation ($\omega_{opt}$). The curvature generated by the optical levitation is then estimated as $\omega_{z}=\sqrt{\omega_{opt}^2-\omega_{magn}^2}$. We find $\omega_{magn}=2\pi\times129(1)$ Hz and $\omega_{opt}=2\pi\times128(1)$ Hz, which are compatible within the error bars. From this we can only deduce an upper bound to $\omega_{z}$ coming from the error bars of the measurement, i.e. $|\omega_{z}|<2\pi\times16$ Hz. In the second test, we compare the evolution of a Bose-Einstein condensate in $|1,-1\rangle$ at 7.5 $a_0$ in the levitating potential and in free space. In Fig.\ref{RBexp} we report the measured Thomas-Fermi radius of the cloud along $z$, as a function of the expansion time. The two datasets correspond to the expansion in free-fall (purple) and in the levitating potential (orange). We compare them to the theoretical curves corresponding to the expansion of the BEC in free space and in a harmonic confining potential along $z$. When the vertical size $R_y$ increases above $\sim 30 \mu$m, we see a deviation from the free-space expansion, which is compatible with the evolution of the BEC with $\omega_z=2\pi\times12$ Hz. This measurement is mostly sensitive to the curvature at large distances, so that it also provides only an upper bound to the effective confinement experienced by the droplets around $z=0$. Finally, note that the sizes of the droplets we observe are always isotropic along the two measured directions, being $\sigma_x$ and $\sigma_z$ compatible within their error bars. Along $x$ the estimated confinement is very weak, $\omega_x=2\pi\times2.2$ Hz. The fact that we do not observe any deformation of the droplet from its predicted spherical geometry let us conclude that the effect of the levitating potential is negligible also on the vertical direction.

\subsection{Absorption imaging of dense clouds}

The droplets we observe have typical densities $n\sim 10^{15}$ cm$^{-3}$. Such high density clouds are difficult to probe with standard absorption imaging, due to saturation effects. In order to image the atoms in the F=1 state of $^{39}$K, we need to use light at two different wavelengths: the first one, called ``repumper'' light, excites the atoms in the $F=2$ state; the second one makes the atoms cycle on the $F = 2 \rightarrow F = 3$ transition, which provides the needed absorption. In order to avoid saturation effects and properly record the density profile of the droplet we use the following technique. We detune the repumper light from resonance by $\sim 100$ MHz. The optical density of the droplet for such detuned laser is smaller than one. In this way we transfer in $F=2$ a cloud with a smaller atom number, but with the same spatial density profile of the droplet. We can then properly measure the size of the cloud, avoiding saturation effects.

The resolution of our imaging system is $\delta_{res}=1.5(1)$ $\mu$m, measured on a non-interacting BEC loaded in a dipole trap with a trapping frequency of 300 Hz. Droplets are typically few $\mu$m large, so we need to take into account the finite imaging resolution to properly estimate their size. The data for $\sigma$ reported in the paper are all deduced by deconvolving the measured size with the imaging resolution as $\sigma=\sqrt{\sigma_{meas}^2-\delta_{res}^2}$.  

During the imaging sequence all magnetic fields need to be off to avoid that the atomic transitions are shifted by the Zeeman effect. The minimum time to switch off the magnetic field $B$ is $\sim 500 \mu$s. During this time, while the field goes to zero, the interatomic scattering lengths change. This could affect the measurement of the size of the cloud, especially when densities are very high, i.e. in the droplet or in the gas phase at short expansion times. We want to estimate this effect, by calibrating it on a mixture whose size can be predicted by a well-established theory, i.e. in the stable mean-field regime. Considering the data reported in Fig.1c of the paper, we can compare the size measured at $t=0$ for the mixture at $\delta a>0$ (orange diamonds) with the theoretical prediction obtained with the coupled Gross-Pitaevskii equations for the double BEC at the same value of $B$. We find that the two results are compatible within the experimental error bars: $\sigma_{meas}=2.4(1) \mu$m and $\sigma_{theor}=2.3\mu$m. We conclude that the switching off of the field has a negligible effect.

\subsection{Three-body losses}

We perform measurements of three-body loss rates on single-component BECs in state 1 and 2 on the magnetic field range probed in the experiment. We analyze the data according to the model in \cite{Zaccanti} and we find $K_{111}/3!=7\times10^{-28}$ cm$^6$/s and $K_{222}/3!=1\times10^{-29}$ cm$^6$/s, with an uncertainty on the absolute values of a factor 2. By performing tests on thermal mixtures, we find that three-body recombination in state 1 is the dominant source of losses. Quantitative estimations for all the different loss rates in the mixture are quite demanding and go beyond the purposes of this work. Detailed measurements would be anyway interesting for future studies, in order to quantitatively characterize the loss dynamics observed in Fig.2b-c of the paper.  

\subsection{Numerical simulations}

The Bose-Bose mixture is described by two coupled Gross-Pitaevskii (GP) equations which include the two-species LHY term discussed in \cite{Petrov}. This is achieved by means of a local density approximation, as recently considered for the description of the formation of macroscopic quantum droplets in dipolar condensates \cite{Ferlaino,Santos,Wachtler}. Namely, the evolution of the system is governed by two equations of the form ($i=1,2$)
\begin{equation}
i\hbar\partial_{t}\psi_{i}= \left(-\frac{\hbar^2}{2m}\nabla^{2}+ V_{ext} +\mu_{i}(n_{1},n_{2}) \right)\psi_{i}
\label{eq:gpe}
\end{equation}
where $\mu_{i}={\delta E}/{\delta n_{i}}$ is the local equilibrium chemical potential for the species $i$ (see e.g. \cite{Menotti}). We consider the case of equal masses ($m_{1}=m_{2}=m$) and define $g_{ij}=4\pi a_{ij}/m$ the coupling constants and $n_i$ the density of species $i$. 
In the present case, the total energy $E$ consists of two contributions: the usual mean-field GP term
\begin{equation}
E_{GP}[n]=\int_{V} \left[\frac12 g_{11} n_{1}^{2} + g_{12}n_{1}n_{2}+ \frac12  g_{22} n_{2}^{2}\right]
\end{equation}
and the LHY contribution \cite{Petrov}
\begin{multline}
E_{LHY}=\frac{8m^{3/2}}{15\pi^{2}\hbar^{3}}\int_{V} \sum_{\pm}\frac{1}{2^{5/2}} \Biggl(g_{11}n_{1}+g_{22}n_{2} \\  \pm \sqrt{(g_{11}n_{1}-g_{22}n_{2})^{2} + 4g_{12}^{2}n_{1}n_{2}}\Biggr)^{5/2}.
\label{eq:elhy2}
\end{multline}
When the two species are prepared with the same density distribution, modulo a scale factor $\alpha$ (corresponding to different atom numbers, namely $n_{2}=\alpha n_{1}\equiv\alpha n$) and the system is close to the onset of the mean-field collapse regime, $g_{12} + \sqrt{g_{11}g_{22}}<0$ (with $g_{ii}>0$), one can further simplify the above expression by setting $g_{12}^{2}=g_{11}g_{22}$ and neglecting small finite-$\delta g$ corrections \cite{Petrov}
\begin{equation}
E_{LHY}=\frac{8m^{3/2}}{15\pi^{2}\hbar^{3}}\int_{V} \left(g_{11}n_{1}+g_{22}n_{2}\right)^{5/2}.
\label{eq:elhy3}
\end{equation}
Eventually, one has ($i,j=1,2$, $i\neq j$)
\begin{equation}
\mu_{i}=\frac{\delta E}{\delta n_{i}}=g_{ii}n_{i} + g_{12}n_{j} + \frac{\delta E_{LHY}}{\delta n_{i}},
\label{eq:mu}
\end{equation}
with
\begin{equation}
\frac{\delta E_{LHY}}{\delta n_{i}}=\frac{32}{3\sqrt{\pi}}g_{ii}\left(a_{11}n_{1}+a_{22}n_{2}\right)^{3/2}.
\label{eq:mulhy}
\end{equation}

Then, Eq. (\ref{eq:gpe}) (with (\ref{eq:mu}), (\ref{eq:mulhy})) is solved by initially loading the condensate in state 2 in the ground state of the external trapping $V_{ext}$, corresponding to the crossed dipole trap of the experiment (by means of a standard imaginary time evolution \cite{dalfovo2}), and then transferring half of the population to state 1, at $t=0$. The subsequent evolution is obtained by solving Eq. (\ref{eq:gpe}) by means of a split-step method which uses fast Fourier transforms (see e.g. \cite{Jackson}).

\end{document}